\documentclass[preprint,showpacs,amsmath,amssymb]{revtex4}
\usepackage{graphicx}
\usepackage{amsfonts}
\usepackage{dcolumn}
\usepackage{bm}

\usepackage{CJK}

\begin{document}
\title{Discontinuous phase transition in an annealed multi-state majority-vote model}

\author{Guofeng Li$^1$}

\author{Hanshuang Chen$^{1}$}\email{chenhshf@ahu.edu.cn}

\author{Feng Huang$^{2}$}

\author{Chuansheng Shen$^{3,4}$}

\affiliation{$^{1}$School of Physics and Materials Science, Anhui
University, Hefei, 230601, China\\
$^2$School of Mathematics and Physics, Anhui Jianzhu University, Hefei 230601, China\\
$^3$Department of Physics, Humboldt University, 12489 Berlin, Germany \\
$^4$Department of Physics, Anqing Normal University, Anqing, 246011,
China }

\date{\today}

\begin{abstract}
In this paper, we generalize the original majority-vote (MV) model
with noise from two states to arbitrary $q$ states, where $q$ is an
integer no less than two. The main emphasis is paid to the
comparison on the nature of phase transitions between the two-state
MV (MV2) model and the three-state MV (MV3) model. By extensive
Monte Carlo simulation and mean-field analysis, we find that the MV3
model undergoes a discontinuous order-disorder phase transition, in
contrast to a continuous phase transition in the MV2 model. A
central feature of such a discontinuous transition is a strong
hysteresis behavior as noise intensity goes forward and backward.
Within the hysteresis region, the disordered phase and ordered phase
are coexisting.

\end{abstract}
\pacs{89.75.Hc, 05.45.-a, 64.60.Cn} \maketitle

\section{Introduction}
Spin models like Ising model play a fundamental role in studying
phase transitions and critical phenomena in the field of statistical
physics and many other disciplines \cite{AJP08000470}. They have
also significant implications for understanding social phenomena
where co-ordination dynamics is observed, e.g., in consensus
formation and adoption of innovations \cite{RMP09000591}. The spin
orientations can represent the choices made by an agent on the basis
of information about its local neighborhood.

The two-state majority-vote (MV2) model with noise, proposed by
Oliveira in 1992 \cite{JSP1992}, is one of the simplest
nonequilibrium generalizations of the Ising model. The model
presents up-down symmetry and a continuous order-disorder phase
transition at a critical value of noise. Studies on regular lattices
showed the critical exponents are the same as those of the Ising
model
\cite{JSP1992,PhysRevE.75.061110,PhysRevE.81.011133,PhysRevE.89.052109,PhysRevE.86.041123},
in accordance with the conjecture by Grinstein \emph{et al.}
\cite{PhysRevLett.55.2527}. The MV2 model has also been extensively
studied in complex networks, including random graphs
\cite{PhysRevE.71.016123,PA2008}, small world networks
\cite{PhysRevE.67.026104,IJMPC2007,PA2015}, scale-free networks
\cite{IJMPC2006(1),IJMPC2006(2),PhysRevE.91.022816}, and some others
\cite{PhysRevE.77.051122,PA2011}. These studies have shown that MV2
models defined on different complex networks belong to different
universality classes and the critical exponents depend on the
topology of the network topologies. The generalization to a
three-state majority-vote model (MV3) on a regular lattice was
considered by \cite{PhysRevE.60.3666,JPA2002}, where the authors
found the critical exponents for this non-equilibrium MV3 model are
in agreement with the ones for the equilibrium three-states Potts
model \cite{RevModPhys.54.235}, supporting again the conjecture of
\cite{PhysRevLett.55.2527}. Recently, Melo \emph{et al.} studied the
MV3 model on random graphs and they found that the critical noise is
a function of the mean connectivity of the graph \cite{JSM2010}. In
\cite{PA2012}, Lima introduced a new zero-state to MV2 model in
square lattices and found that this model also falls into the Ising
universality. Costa \emph{et al.} generalized the state variable of
the MV model from discrete case to continuous one, and they found
that a Kosterlitz-Thouless-like phase appears in low values of noise
\cite{PhysRevE.71.056124}.

In the present work, we generalize the MV model to arbitrary
multiple states where each agent is allowed to be in one of $q$
different states with $q$ being an integer no less than two. In each
step, an agent is randomly chosen and assumes that its state aligns
with the state of the majority of its neighborhood with probability
$1-f$ and with the minority state with probability $f$. Herein, the
number of neighboring spins of any agent is fixed that randomly
selected from all the other agents at each step. We mainly focus on
the difference in the nature of the order-disorder phase transition
between the MV2 model and the MV3 model. Interestingly, we find that
in the MV3 model there exists a strong hysteresis behavior as noise
intensity goes forward and backward, the main feature of a
discontinuous phase transition, in contrast to a continuous phase
transition in the MV2 model. A mean-field theory is used to explain
the nature of this phase transition.

\section{Model}
Here, we generalize the original MV model from two states to
arbitrary multiple states. The model is defined as follows. Each
agent $i$ ($= 1, \ldots ,N$) can be in any one of $q$ states:
${\sigma _i} \in \{ 1, \cdots ,q\}$. The number of the neighboring
agents of agent $i$ in any state $\alpha$ can be calculated as
${n_\alpha } = \sum\nolimits_{j \in \mathcal {N}(i)} {{\delta
_{\alpha ,{\sigma _j}}}}$, where the summation takes over all
neighboring nodes of $i$. ${\delta _{\alpha ,\beta }}$ is Kronecker
symbol defined as ${\delta _{\alpha ,\beta }} = 1$ if $\alpha=\beta$
and ${\delta _{\alpha ,\beta }} = 0$ otherwise. With the probability
$1-f$, the agent $i$ take the same value as the majority spin, i.e.,
${\sigma _i} = {\left. \alpha \right|_{{n_\alpha } = \max \{ {n_1},
\cdots ,{n_q}\} }}$. With the supplementary probability $f$ the
agent $i$ takes the same value as the minority spin, i.e., ${\sigma
_i} = {\left. \alpha \right|_{{n_\alpha } = \min \{ {n_1}, \cdots
,{n_q}\} }}$. If there are more than two possible states in the
majority spin or in the minority spin, we randomly choose one of
them. If $q=2$, we recover the original MV2 model. In the present
work, we consider the model defined on random degree-regular
networks with annealed randomness, i.e., the number of neighbors of
each agent is fixed but they are randomly chosen from all others at
each time step.

To characterize the critical behavior of the model, we define the
order parameter as
\begin{eqnarray}
 m = \left| {{N^{ - 1}}\sum\nolimits_{j = 1}^N {{e^{{{I2\pi {\sigma _j}} \mathord{\left/
 {\vphantom {{I2\pi {\sigma _j}} q}} \right.
 \kern-\nulldelimiterspace} q}}}} } \right|\label{eq1}
\end{eqnarray}
where $I$ is the imaginary unit. $m>0$ corresponds to an ordered
state, and $m=0$ to a disordered state. Furthermore, we define the
susceptibility $\chi$ and Binder's fourth-order cumulant $U$ as
\begin{eqnarray}
\chi  = N\left[ {\left\langle {{m^2}} \right\rangle  -
{{\left\langle m \right\rangle }^2}} \right]\label{eq2}
\end{eqnarray}
\begin{eqnarray}
U = 1 - \frac{{\left\langle {{m^4}} \right\rangle
}}{{3{{\left\langle {{m^2}} \right\rangle }^2}}}\label{eq3}
\end{eqnarray}
where $\left\langle  \cdots  \right\rangle$ denotes time averages
taken in the stationary regime.

\section{Simulation results}
We have performed extensive Monte Carlo (MC) simulation with various
systems of size $N=500$, $1000$, $2000$, $4000$, $8000$, $16000$. At
each elementary step, an agent is randomly chosen and tried to
update its state according to its neighboring agents as defined
before. Here, the neighboring agents are selected randomly from all
other agents and the number of neighboring agents is fixed at $z=4$.
On each MC step (MCS), each agent is tried to update its state once
on average. To obtain steady values of $m$, the first $10^5$ MCS are
discarded and the following $10^5$ MCS are used to calculate time
averages of $m$. At the critical region, larger runs are performed
with $2\times10^5$ MCS to reach the steady state and $10^6$ for
computing time averages.

\begin{figure}
\centerline{\includegraphics*[width=1.0\columnwidth]{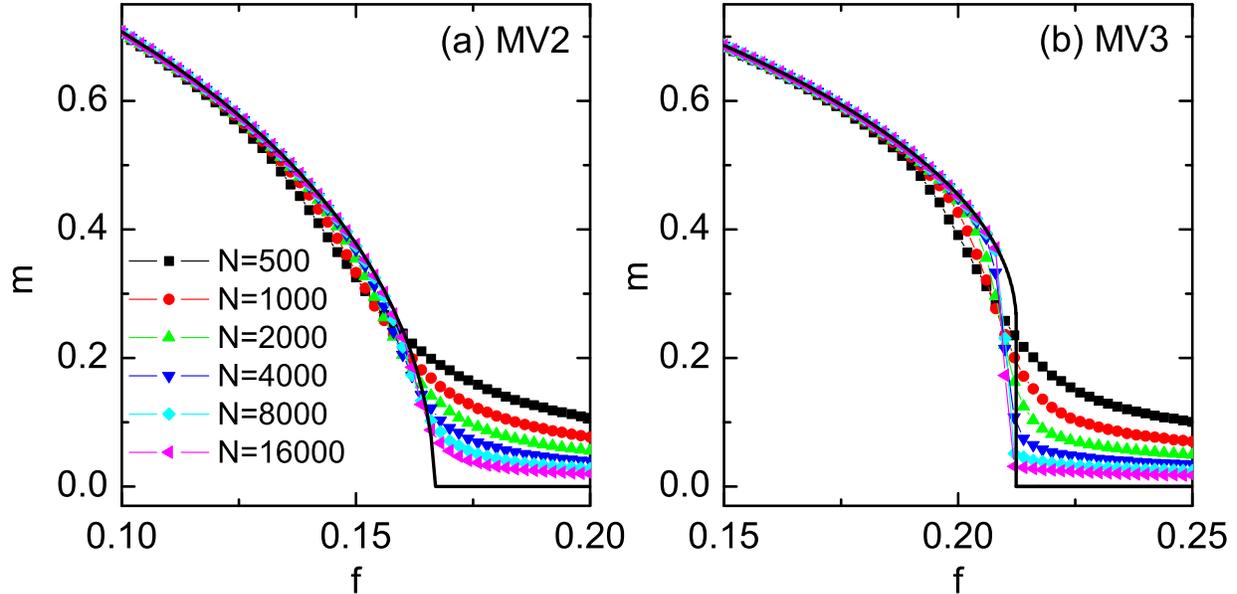}}
\caption{(color online). Dependence of the order parameter $m$ on
the noise intensity $f$ in the MV2 model (a) and in the MV3 model
(b). The lines correspond to mean-field results. \label{fig1}}
\end{figure}

\begin{figure}
\centerline{\includegraphics*[width=1.0\columnwidth]{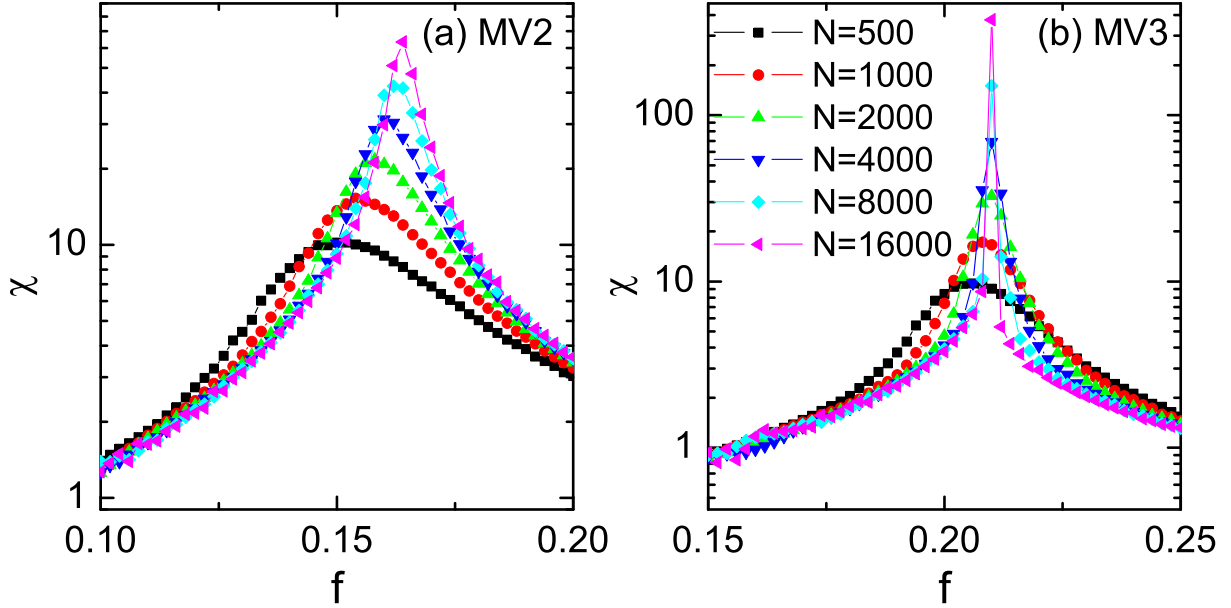}}
\caption{(color online). Dependence of the susceptibility $\chi$ on
the noise intensity $f$ in the MV2 model (a) and in the MV3 model
(b). \label{fig2}}
\end{figure}

In Fig.\ref{fig1}(a) and Fig.\ref{fig1}(b), we show $m$ as a
function of the noise intensity $f$ in the MV2 model and in the MV3
model, respectively. For the MV2 model, one can clearly see that the
model displays a continuous order-disorder phase transition as the
noise intensity $f$ increases. The maximum in the susceptibility
$\chi$ indicates that there exists an effective transition noise
$f_c(N)$ for a given $N$, as shown in Fig.\ref{fig2}(a). $f_c(N)$
increases with $N$ and such a scaling determines a critical exponent
of the continuous phase transition. For the MV3 model, especially
for a lager $N$, we find that $m$ displays a sharp change close
order-disorder phase transition. On the other hand, as shown in
Fig.\ref{fig2}(b), The susceptibility $\chi$ also shows a maximum at
$f_c(N)$ but the change of $\chi$ near $f_c(N)$ becomes sharper and
the values of $\chi$ at $f_c(N)$ becomes larger. These results may
suggest the nature of phase transition in the MV3 model is different
from that of the MV2 model.

\begin{figure}
\centerline{\includegraphics*[width=1.0\columnwidth]{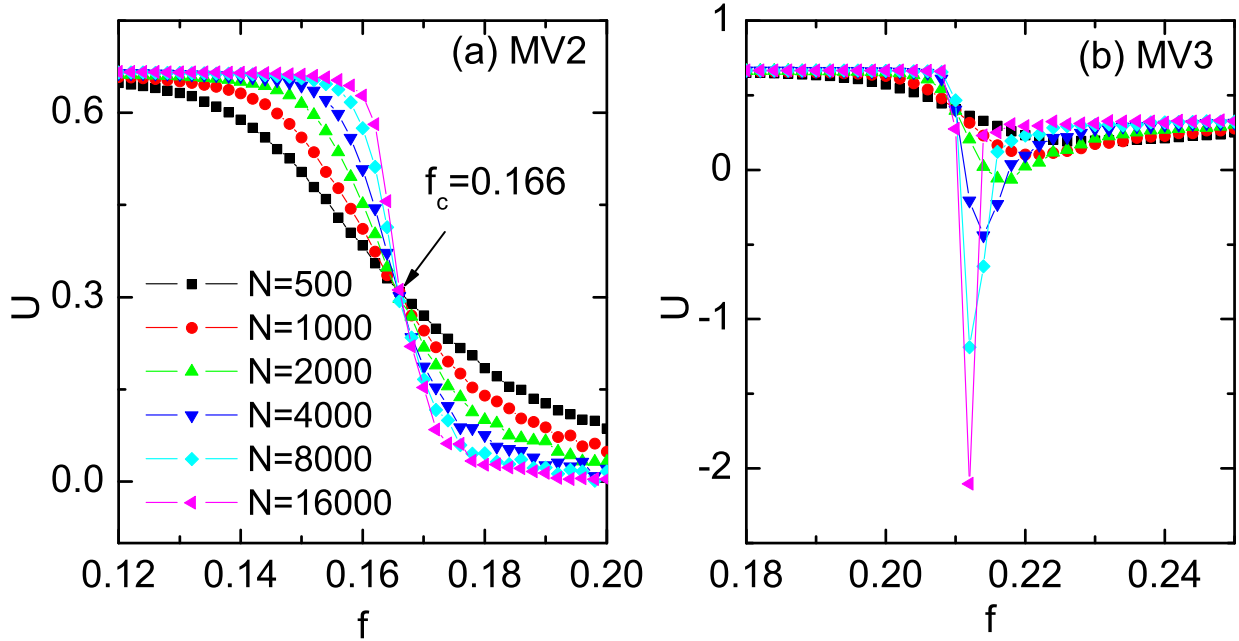}}
\caption{(color online). Dependence of the Binder's fourth-order
cumulant $U$ on the noise intensity $f$ in the MV2 model (a) and in
the MV3 model (b). \label{fig3}}
\end{figure}

To verify our hypothesis, we compare the results of the Binder's
cumulant $U$ in the MV2 model and in the MV3 model. For the MV2
model, the phase transition is continuous and the cumulants for
different $N$ intercept each other at $f=f_c^{MV2}$, providing a
convenient estimate for the value of the critical noise
$f_c^{MV2}\simeq0.166$, as shown in Fig.\ref{fig3}(a). In
Fig.\ref{fig3}(b), we plot the curves $U \sim f$ for different $N$
in the MV3 model. In contrast to the case in MV2 model, $U$ exhibits
a nonmonotonic dependence on $f$ and there exists a minimum near the
onset of phase transition. For larger $N$, $U$ falls to negative
values near phase transition that is the sign of a discontinuous
transition, together with the coexistence of the ordered and
disordered phases expected then \cite{Binder1997}.

\begin{figure}
\centerline{\includegraphics*[width=1.0\columnwidth]{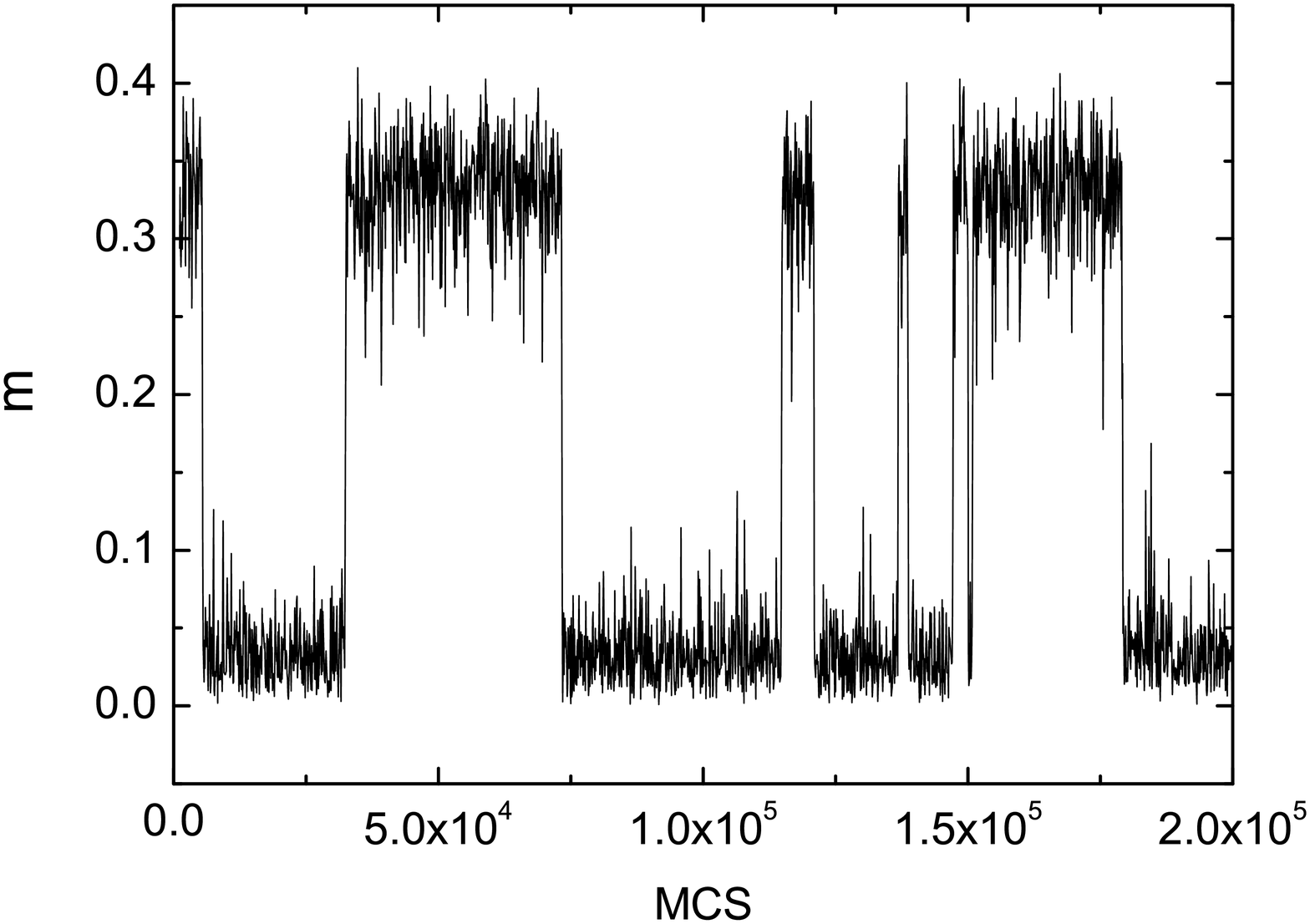}}
\caption{A long time series of order parameter $m$ close the phase
transition in the MV3 model. The parameters are $f=0.21$ and
$N=16,000$. \label{fig4}}
\end{figure}

\begin{figure}
\centerline{\includegraphics*[width=1.0\columnwidth]{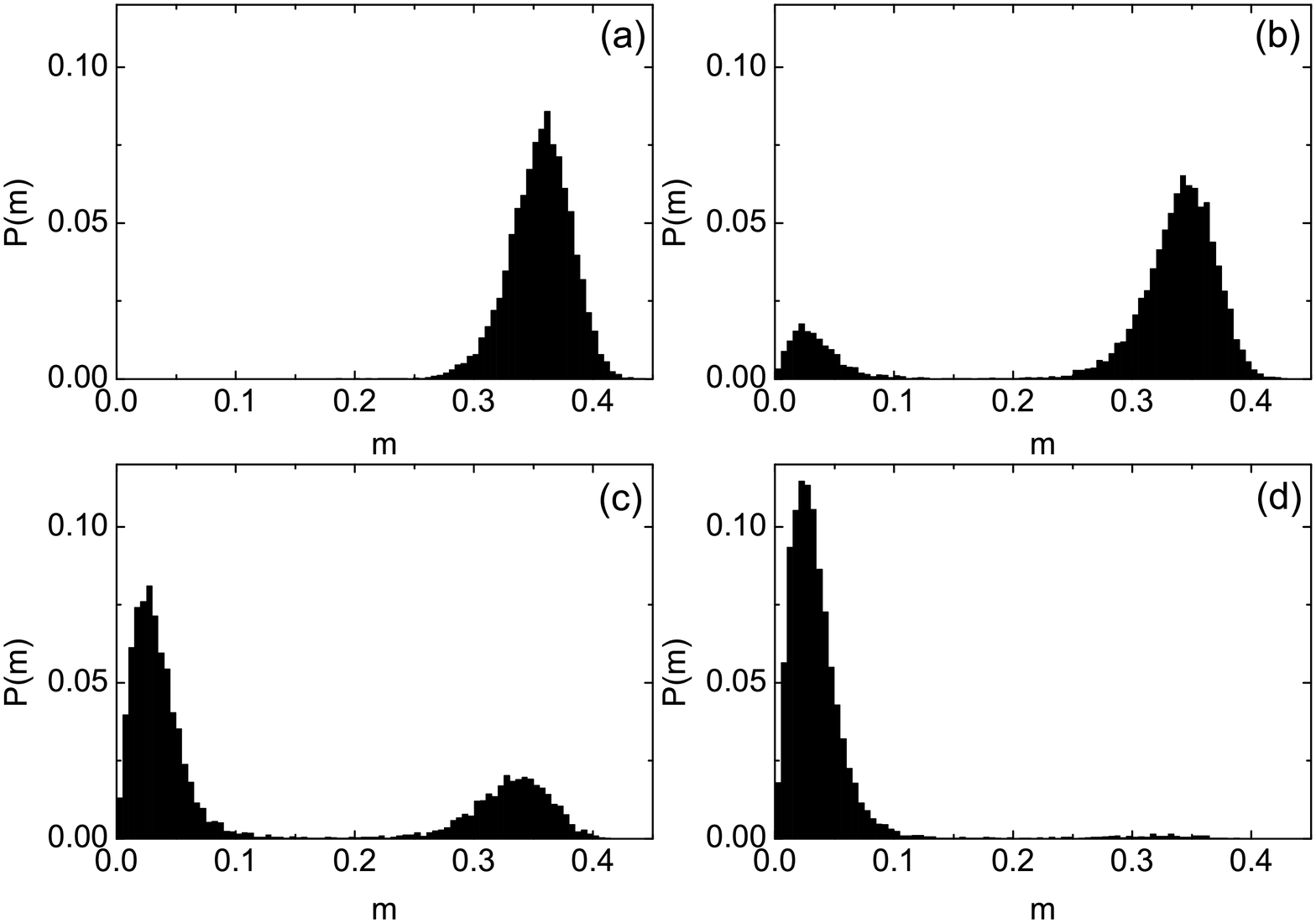}}
\caption{Histogram of the values of order parameter $m$ across a
discontinuous phase transition in the MV3 model. The noise intensity
$f$ are $0.209$, $0.2098$, $0.2102$, and $0.211$ from (a) to (d),
respectively. The size of the system is $N=16,000$. \label{fig5}}
\end{figure}

To further validate the discontinuous nature of phase transition in
the MV3 model, in Fig.\ref{fig4} we plot a long time series of $m$
close the transition noise for $N=16,000$. It is clearly observed
that the system flips between the ordered phase and disordered phase
due to the finite-size fluctuations. This implies that at this noise
the system is in the coexistence of the ordered and disordered
phases, that is a typical of discontinuous phase transition. In
Fig.\ref{fig5}, we show the distribution of the values of $m$ for
several distinct values of $f$ across the phase transition in the
MV3 model for $N=16,000$. For $f=0.209$ (Fig.\ref{fig5}(a)), $m$ has
a unique maximum around $m=0.36$, which corresponds to an ordered
phase. For $f=0.2098$ (Fig.\ref{fig5}(b)), $m$ displays two peaks
around $m=0$ and $m=0.34$ which indicate the coexistence of an
ordered phase and a disordered one. Upon increasing the value of
$f$, the peak at $m=0$ is enhanced, while the other peak is
depressed (see Fig.\ref{fig5}(c) for $f =0.2102$), implying that the
disorder phase becomes more stable. For even larger values of $f$,
there remains a single peak around $m=0$, which indicates that the
system is in a disordered phase (see Fig.\ref{fig5}(d) for
$f=0.211$). This set of histograms further verify that the MV3 model
undergoes a discontinuous order-disorder transition. Note that if
the network configuration among agents is quenched other than
annealed used in the present work, the MV3 also displays a
continuous phase transition as in the MV2 model \cite{JSM2010}.

\section{Mean-field theory}
To analytical understand the difference on the nature of phase
transitions between the MV2 model and the MV3 model, we will present
a mean-field theory to determine steady values of $m$ as a function
of $f$. To the end, let $x_\alpha$ denote the probability that each
node is in the state $\alpha$ ($\alpha=1,\ldots,q$). The time
evolution of $x_\alpha$ can be written as
\begin{eqnarray}
{\dot x_\alpha } = \sum\limits_{\beta  \ne \alpha } {{x_\beta }}
{w_{\beta  \to \alpha }} - {x_\alpha }\sum\limits_{\beta  \ne \alpha
} {{w_{\alpha  \to \beta }}} \label{eq4}
\end{eqnarray}
where $w_{\alpha  \to \beta }$ is the transition rate from state
$\alpha$ to state $\beta$. According to our model, the rate
$w_{\alpha  \to \beta }$ can be expressed as the sum of two parts,
\begin{eqnarray}
{w_{\alpha  \to \beta }} = (1 - f){P_\beta } + f{{\tilde P}_\beta }
\label{eq5}
\end{eqnarray}
where the first part is the probability of taking majority-rule
$1-f$ by the probability $P_\beta$ that the state $\beta$ is in the
majority state, and the second part is the probability of taking
minority-rule $f$ by the probability $\tilde P_\beta$ that the state
$\beta$ is in the minority state. Using the normalization
conditions, $\sum\nolimits_\beta {{x_\beta }}=\sum\nolimits_\beta
{{P_\beta }} = \sum\nolimits_\beta {{{\tilde P}_\beta }}  = 1$,
Eq.(\ref{eq4}) can be simplified to
\begin{eqnarray}
{{\dot x}_\alpha } =  - {x_\alpha } + (1 - f){P_\alpha } + f{{\tilde
P}_\alpha }. \label{eq6}
\end{eqnarray}
In the steady state $\dot x_\alpha=0$, we have
\begin{eqnarray}
{x_\alpha } = (1 - f){P_\alpha } + f{{\tilde P}_\alpha }.
\label{eq7}
\end{eqnarray}
Let $n_\alpha$ denote the number of neighbors in the state $\alpha$
among all $z$ neighbors, and the probability of a given
configuration $\{n_\alpha\}$ can be written as multinominal
distribution,
\begin{eqnarray}
p(\{ {n_\alpha }\} ) = \frac{{z!}}{{\prod\limits_\alpha  {{n_\alpha
}!} }}\prod\limits_\alpha  {x_\alpha ^{{n_\alpha }}}. \label{eq8}
\end{eqnarray}
Thus, $P_\alpha$ and $\tilde P_\alpha$ can be expressed as
\begin{eqnarray}
{P_\alpha } = \sum\limits_{\{ {n_\alpha }\} |{n_\alpha } > {n_\beta
},\forall \beta  \ne \alpha } {p(\{ {n_\alpha }\} )}  +
\sum\limits_{\{ {n_\alpha }\} |{n_\alpha } = {n_\beta } > {n_\gamma
},\forall \gamma  \ne \alpha ,\beta } {\frac{1}{2}p(\{ {n_\alpha }\}
)}  +  \cdots,\label{eq9}
\end{eqnarray}
and
\begin{eqnarray}
{\tilde P_\alpha } = \sum\limits_{\{ {n_\alpha }\} |{n_\alpha } <
{n_\beta },\forall \beta  \ne \alpha } {p(\{ {n_\alpha }\} )}  +
\sum\limits_{\{ {n_\alpha }\} |{n_\alpha } = {n_\beta } < {n_\gamma
},\forall \gamma  \ne \alpha ,\beta } {\frac{1}{2}p(\{ {n_\alpha }\}
)}  +  \cdots,\label{eq10}
\end{eqnarray}

\begin{figure}
\centerline{\includegraphics*[width=1.0\columnwidth]{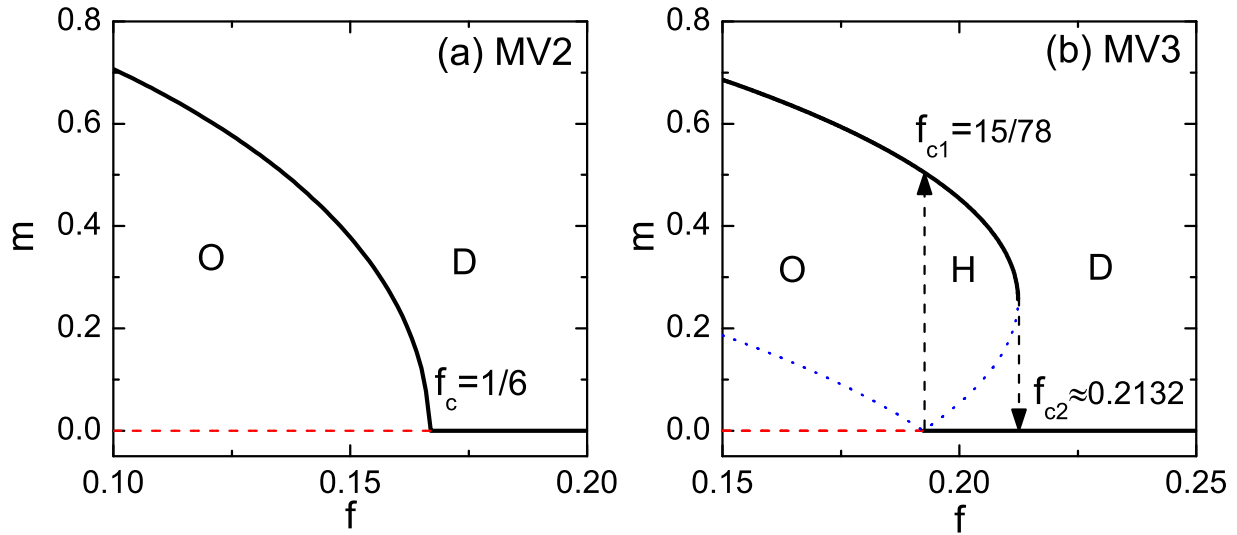}}
\caption{(color online). The steady solutions of $m$ in the MV2
model (a) and in the MV3 model (b). The MV2 model displays a
continuous order-disorder (O-D) phase transition at
$f=f_c^{MV2}=\frac{1}{6}$. The MV3 model displays a discontinuous
O-D phase transition at $f=f_{c_2}^{MV3}\simeq0.2132$ and a
discontinuous D-O phase transition at
$f=f_{c_1}^{MV3}=\frac{15}{78}$ as $f$ varies forward and backward.
In the hysteresis (H) region between $f_{c_1}^{MV3}$ and
$f_{c_2}^{MV3}$, the MV3 model is in the coexistence of O phase and
D phase. The solid line, dashed line and dotted line correspond that
the solutions are stable, unstable and saddle, respectively.
\label{fig6}}
\end{figure}

In the present work, the number of neighbors of each agent is kept
at $z=4$. For the MV2 model, $P_\alpha$ and $\tilde P_\alpha$ are
\begin{eqnarray}
\left\{ \begin{gathered}
  {P_\alpha } = x_\alpha ^4 + 4x_\alpha ^3{x_\beta } + 3x_\alpha ^2x_\beta ^2 \hfill \\
  {{\tilde P}_\alpha } = x_\beta ^4 + 4{x_\alpha }x_\beta ^3 + 3x_\alpha ^2x_\beta ^2 \hfill \\
\end{gathered}  \right.,\label{eq11}
\end{eqnarray}
where $\alpha ,\beta  \in \{ 1,2\}$ and $\alpha  \ne \beta$. One can
easily check that ${x_\alpha }= \frac{1}{2}$ ($\alpha  = 1,2$) is
always a set of solution of Eq.(\ref{eq4}) for any given $f$. This
trivial solution corresponds to a disordered phase. The solution
loses its stability when the maximal eigenvalue of Jacobi matrix is
larger than zero, which leads to the analytical value of the
critical noise $f_c^{MV2}=\frac{1}{6}$. The other solutions can be
obtained by numerically solving Eq.(\ref{eq7}) combined with
Eq.(\ref{eq11}). In Fig.\ref{fig6}(a), we show the steady solutions
of $m$ as a function of $f$. The solid and dashed lines indicate the
stable and the unstable steady solutions, respectively. The MV2
model presents a continuous order-disorder phase transition at
$f=f_c^{MV2}=\frac{1}{6}$.

For MV3 model, $P_\alpha$ and $\tilde P_\alpha$ are
\begin{eqnarray}
\left\{ \begin{gathered}
  {P_\alpha } = x_\alpha ^4 + 4x_\alpha ^3{x_\beta } + 4x_\alpha ^3{x_\gamma } + 3x_\alpha ^2x_\beta ^2 + 3x_\alpha ^2x_\gamma ^2 + 12x_\alpha ^2{x_\beta }{x_\gamma } \hfill \\
  {{\tilde P}_\alpha } = \frac{1}{2}x_\beta ^4 + \frac{1}{2}x_\gamma ^4 + 4x_\beta ^3{x_\gamma } + 4{x_\beta }x_\gamma ^3 + 6x_\beta ^2x_\gamma ^2 + 6{x_\alpha }x_\beta ^2{x_\gamma } + 6{x_\alpha }{x_\beta }x_\gamma ^2 \hfill \\
\end{gathered}  \right.\label{eq12}
\end{eqnarray}
where $\alpha ,\beta ,\gamma  \in \{ 1,2,3\}$ and $\alpha  \ne \beta
\ne \gamma$. Since $P_\alpha=\tilde P_\alpha=\frac{1}{3}$ at
${x_\alpha }= \frac{1}{3}$ ($\alpha  = 1,2,3$) (corresponding to a
disordered phase), one can easily check that Eq.(\ref{eq4}) holds
for any given $f$. As mentioned before, this trivial solution loses
its stability when the maximal eigenvalue of Jacobi matrix is larger
than zero. It immediately leads to the analytical value of
$f_{c_1}^{MV3}=\frac{15}{78}\simeq0.1923$. The other solutions can
be obtained by numerically solving Eq.(\ref{eq7}) combining with
Eq.(\ref{eq12}). In Fig.\ref{fig6}(b), we show the steady solutions
of $m$ as a function of $f$. The solid line, dashed line and dotted
line correspond that the solutions are stable, unstable and saddle,
respectively. As $f$ varies forward and backward, $m$ presents two
discontinuous transitions at $f=f_{c_2}^{MV3}\simeq0.2132$ and at
$f=f_{c_1}^{MV3}$. This indicates that there exists a hysteresis
region at $f_{c_1}^{MV3}<f<f_{c_2}^{MV3}$ for which the ordered
phase and disordered phase are coexisting.

\section{Conclusions and Discussion}
In conclusion, we have generalized the original MV model from two
states to arbitrary multiple states. We have made the detailed
comparison on the nature of phase transitions between the MV2 model
and the MV3 model defined on annealed networks. Interestingly, we
have found that the discontinuous nature of order-disorder phase
transition in the MV3 model, qualitatively different from the
continuous phase transition in the MV2 model. By a mean-field
analysis, we have obtained the steady solutions of order parameter
$m$ as a function of noise $f$. Based on the stability analysis of
the solutions, we have obtained the analytical value $f_{c_2}^{MV3}$
of noise for the onset of the discontinuous disorder-order phase
transition. Another value $f_{c_1}^{MV3}$ of noise for the onset of
the discontinuous order-disorder phase transition is also
numerically estimated. In the hysteresis region between these two
transition noises, the MV3 model is coexistent of an ordered phase
and a disordered one. For a finite size system, the fluctuation can
drive the flips between these two phases. Our findings may suggest a
novel mechanism of first-order phase transitions in nonequilibrium
systems. At last, we should emphasize that there exists the
essential difference in the nature of phase transition between the
annealed and quenched MV3 models \cite{JSM2010}.

\begin{acknowledgments}
This work was supported by National Science Foundation of China
(Grants Nos. 11205002, 61473001, 11475003, 11405001), the Key
Scientific Research Fund of Anhui Provincial Education Department
(Grant No. KJ2016A015) and ``211" Project of Anhui University (Grant
No. J01005106).
\end{acknowledgments}

%
\bibliographystyle{apsrev}

\end{document}